\documentclass[aps,prb,preprint,groupedaddress,superscriptaddress]{revtex4}
\usepackage{graphicx}
\usepackage{epstopdf}
\usepackage{amsmath}
\usepackage{amssymb}
\begin{document}
\title{On-chip phononic time lens}

\author{M. Kurosu}
\email{kurosu\_megumi\_s5@lab.ntt.co.jp}
\affiliation{NTT Basic Research Laboratories, NTT Corporation, Atsugi-shi, Kanagawa 243-0198, Japan}
\affiliation{Department of Physics, Tohoku University, Sendai 980-8578, Japan}

\author{D. Hatanaka}
\affiliation{NTT Basic Research Laboratories, NTT Corporation, Atsugi-shi, Kanagawa 243-0198, Japan}
\author{K. Onomitsu}
\affiliation{NTT Basic Research Laboratories, NTT Corporation, Atsugi-shi, Kanagawa 243-0198, Japan}
\author{H.Yamaguchi}

\affiliation{NTT Basic Research Laboratories, NTT Corporation, Atsugi-shi, Kanagawa 243-0198, Japan}
\affiliation{Department of Physics, Tohoku University, Sendai 980-8578, Japan}

\begin{abstract}
{\bf
The ability to manipulate phonon waveforms in continuous media has attracted significant research interest and is crucial for practical applications ranging from biological imaging to material characterization.
Although several spatial focusing techniques have been developed, these systems require sophisticated artificial structures, which limit their practical applications. This is because the spatial control of acoustic phonon waves is not as straightforward as photonics so there is a strong demand for an alternative approach. 
Here we demonstrate a phononic time lens in a dispersive one-dimensional phononic crystal waveguide, which enables the temporal control of phonon wave propagation. 
Pulse focusing is realized at a desired time and position with chirped input pulses that agree perfectly with the theoretical prediction.
This technique can be applied to arbitrary systems and will offer both an improvement in time and spatial sensing resolution and allow the creation of a highly intense strain field, enabling the investigation of novel nonlinear phononic phenomena such as phononic solitons and rogue waves.
}
\end{abstract}

\maketitle

Acoustic phonons have been widely utilized in various applications and are especially important as a tool for nonintrusive sensing in such areas as biological imaging and defence systems\cite{Buckingham_Imaging_1992, Monstafa_Ultrasound_1998, Kushibiki_Material_1985,796320}. The capabilities of these phonon systems, for example imaging resolution and accuracy, are determined by the spatial size and energy density of generated waves.  
To improve these capabilities, several spatial phonon focusing techniques have been proposed where engineered structures such as the phased arrays, negative-index materials and acoustic metamaterials enable acoustic phonons to be focused in tiny spatial regions\cite{ PhysRevLett.93.024301, Spadoni_Generation_2010, Chen_Enhanced_2014}. However, these conventional focusing techniques, which use a spatial lens, rely on sophisticated spatially designed structures, and this limits their practical use.

A time lens, which is a temporal analogue to a spatial lens, has been introduced in the field of optics \cite{Akhmanov_1968,Treacy_1969, foster2008silicon-chip-based, Salem_2013}.
According to space-time duality, which describes the mathematical equivalence between paraxial-beam diffraction and dispersive pulse broadening, the effect of dispersion in a medium is the key to realizing a time lens. Importantly, this temporal focusing method can be applied in an arbitrary dispersive material in which it imparts quadratic time-varying phase shift.
On the basis of this concept, various useful ideas have been developed including temporal imaging, temporal magnification and real-time spectroscopy in optics \cite{Salem_Optical_2008,Bennett_Temporal_1994, Kolner_1989, Solli_2007}.
By introducing the approach in the field of phononics, we have realized a phononic time lens in a one-dimensional (1D) phononic crystal waveguide (PnC WG)\cite{Hatanaka_Phonon_2015, Hatanaka_Phonon_2014} which is constructed by nanoelectromechanical systems (NEMS) technology. Ultrasound phonon waves travelling through the WG experience pulse broadening due to the group velocity dispersion (GVD) effect. 
This GVD effect can be used to compensate for the frequency modulation of the initial pulse that leads to temporal focusing. 
By further incorporating excellent mechanical properties of NEMS such as high-quality factors, integrability and nonlinearity into this device\cite{singh2014optomechanical, Weber_Coupling_2014, okamoto2013coherent,faust2013coherent}, the ability to temporally focus the travelling phonons will open up the possibility of developing an ultrashort and highly intense phonon pulse generator, which will be useful for practical applications as in the case of optical laser systems\cite{RevModPhys.78.309, Perry917, gauthier2016chirped}, and investigaing nonlinear phononic phenomena.

The PnC WG consists of a 1 mm long membrane made from a GaAs/AlGaAs heterostructure as shown in Fig. 1(a), where periodically-arrayed air holes with a pitch of 8 $\mu$m are formed along the WG that can be used to suspend the membrane with a width of 22 $\mu$m by selectively etching the Al$_{0.65}$Ga$_{0.35}$As layer.
The application of an alternating voltage to an electrode located at both edges of the WG induces phonon vibrations due to the piezoelectric effect.
The resultant vibrations travel down the WG and are detected in an optical interferometer. 

Figure 1(b) shows the experimental transmission spectrum of the device and the corresponding band structure calculated by using a finite element method (FEM) simulation (COMSOL Multiphysics).
The phonon vibrations are observed in 3.5-7.5 MHz owing to the presence of 1st phonon band, where the phonon vibrations propagate in the WG and are reflected at both clamping edges thus resulting in the generation of equidistant Fabry-Perot peaks in the spectral response.
On the other hand, phonon waves around 8 MHz experience Bragg reflection from the periodic air holes, giving rise to a phonon bandgap, and this prevents them from propagating in the WG\cite{Hatanaka_Phonon_2015, Hatanaka_Phonon_2014}. Above the bandgap, there is a new phonon branch that again allows the phonon vibrations to be guided. In the experiments described below, we focus on the 1st phonon branch to investigate the temporal dynamics of the phonon vibrations in the device.

A transverse deflection $z(x, t)$ travelling in a 1D PnC WG can be described by Euler-Bernoulli equations as\cite{Nayfeh_book},
\begin{equation}
EI\frac{\partial^4z(x,t)}{\partial x^4}+\rho S\frac{\partial^2z(x,t)}{\partial t^2}+\alpha_1z(x,t)+\alpha_3z^3(x,t)=0
\label{eq:w}
\end{equation}
where $E$ is the Young's modulus, $S$ and $I$ are the area and moment of inertia of the cross section, $\rho$ is the density of the WG  per unit length and $\alpha_1$ and $\alpha_3$ are the elastic coefficients of the WG. 
To solve equation (\ref{eq:w}), the slowly varying amplitude of travelling wave is assumed that the envelope of a travelling vibration pulse centred around wavenumber $k$ and angular frequency $\omega$ varies slowly in temporal and spatial domain on the moving-frame with group velocity,
\begin{equation}
z(x,t) = A(x,t){\rm exp}\bigl[ i(kx-\omega t)\bigr].
\label{eq:za}
\end{equation}
We introduce the linear loss term $\eta$, then $A(x, t)$ satisfies the following equation\cite{Nayfeh_book},
\begin{equation}
i\frac{\partial A}{\partial x}=-\frac{i\eta}{2}A+\frac{k_2}{2}\frac{\partial^2 A}{\partial t^2}-ik_1\frac{\partial A}{\partial t}-\xi A^2 {\bar A}, 
\label{eq:aa}
\end{equation}
where $k_1=\frac{\partial k}{\partial \omega}\equiv v^{-1}_g$ is the inverse of the group velocity, $k_2=\frac{\partial^2 k}{\partial \omega^2}$ is the GVD coefficient and $\xi$ is a nonlinear parameter. These terms are determined by the material and geometric parameters of the WG in equation (\ref{eq:w}). Thus, the dynamics of phonon propagation based on the Euler-Bernoulli equation corresponds to a nonlinear Schr$\rm \ddot{o}$dinger equation (NLSE), which is used to describe the dynamics of an optical wave propagating in a dispersive medium\cite{Agrawal_5th, Mollenauer_83, Blanco_soliton_2014}. 
By neglecting the nonlinear term in equation (\ref{eq:aa}) and using the new time coordinate $T$ moving with the group velocity $v_g$,
\begin{equation}
T=t-\frac{x}{v_{\rm g}}=t-k_1x.
\label{eq:www}
\end{equation}
At the same time we introduce a normalized amplitude $U(x,T)$, 
\begin{equation}
A(x, T)=A_0{\rm exp}(-\eta x/2)U(x, T),
\end{equation}
where $A_0$ is the peak amplitude of the input pulse, and the equation can be simplified as
\begin{equation}
{\rm i}\frac{\partial U}{\partial x} = \frac{k_2}{2}\frac{\partial ^2 U}{\partial T^2}.
\label{eq:u}
\end{equation}
Interestingly, equation (\ref{eq:u}) is mathematically equal to the paraxial wave equation that governs the diffraction of CW light. If we assume a Gaussian pulse as an input, the normalized amplitude at distance $x$ is given by\cite{Agrawal_5th},
\begin{equation}
U(x, T) = \frac{T_0}{(T_0^2-{\rm i}k_2x)^{1/2}}{\rm exp}\biggl(-\frac{T^2}{2(T_0^2-{\rm i}k_2x)}\biggr),
\label{eq:uu}
\end{equation}
where $T_0$ is the half-width at the 1/$e$-intensity point, and its output width $T_1$ is written as
\begin{equation}
T_1(x)=T_0\sqrt{1+\Bigl(\frac{|k_2|x}{T_0^2}\Bigr)^2}.
\label{eq:uuu}
\end{equation}
Thus a phonon temporal waveform travelling down a dispersive medium depends on the absolute value of the GVD coefficient that enables the temporal pulse width to be broadened further due to the larger dispersion.

To elucidate the temporal characteristics of this device experimentally, phonon vibrations are measured at the left edge by exciting them with a Gaussian pulse with $T_0 = 1.2\ \mu$s from the right edge as shown in Fig. 2(a). 
This time-of-flight measurement also enables the group velocity $v_{\rm g}$ and the GVD coefficient $k_2$ to be estimated as function of excitation frequency as shown in Figs. 2(b) and 2(c) respectively. 
The experimental $v_{\rm g}$ value increases with increasing frequency i.e. anomalous dispersion ($k_2 < 0$) in the band except near the bandgap where there is a large reduction in $v_{\rm g}$, i.e. normal dispersion ($k_2 > 0$), because of the decreased slope of the band. 
In addition, the temporal waveform of the pulse around the band edges is distorted and becomes asymmetric with an oscillation near the trailing edge caused by a contribution from the 3rd order dispersion effect (see Supplementary Information for more detail). 
These experimental results can be well reproduced by the calculation results obtained with FEM-simulated band structures. 

The temporal pulse widths can also be estimated from the time-of-flight measurement in Fig. 2(a) by fitting a Gaussian envelope to the output waveforms. Figures 3(b) and 3(c) reveal the output pulse widths at various excitation frequencies when exciting Gaussian pulses with $T_0 = 1.2$ and 0.7 $\mu$s respectively and measuring them at distances $x=$ 1, 3 and 5 mm (see Fig. 3(a)).
As theoretically predicted, the pulse widths are greatly increased as the frequency approaches the band edges where $| k_2|$ is large and the GVD-induced pulse broadening becomes distinct with increasing propagation distance as shown in Figs. 3(b)and 3(c)\cite{Marcuse_Pulse_1980}. 
In particular, shortening the input pulse from $T_0=1.2$ $\mu$s to 0.7 $\mu$s, namely spectrally broadening, leads to the significant influence of the propagation distance on the output width as shown in Fig. 3(c). 

In the previous experiment we confirmed that the pulse is broadened during propagation thanks to the GVD effect. In an anomalous (normal) dispersion regime, the high-frequency components of the pulse travel faster (slower) than its low-frequency components, thus allowing the injected un-chirped pulse to be frequency chirped and broadened, where the product of the temporal and spectral widths is not transform limited. 
In general, this effect is unfavorable for efficient phonon guiding. 
However, we utilize the disadvantage to demonstrate temporal focusing, namely the compression and amplification, of the travelling phonon pulse. 
Here, a frequency-chirped pulse is excited as the input, and frequency modulation within the pulse is compensated for by the GVD via propagation, thus resulting in the pulse being compressed to the lower limit for the pulse width determined by a given spectrum, namely transform limited, and the peak amplitude being amplified. 
Thus this temporal analogue of a spatial lens, called a time lens, enables the temporal focusing of a phonon wave\cite{Salem_2013}. The dynamics of the chirped pulse evolution in the WG can also be described by NLSE, and the output pulse width $T_2$ is given by
\begin{equation}
T_2(x) = T_0\sqrt{\Bigl(1+\frac{Ck_2x}{T_0^2}\Bigl)^2+\Bigl(\frac{k_2x}{T_0^2}\Bigl)^2}.
\label{eq:t}
\end{equation}
$C$ is a chirp parameter that is positive (negative) when the frequency increases (decreases) linearly from the leading to the trailing edge, and is defined by $\Delta f=\sqrt{(1+C^2)}/2\pi T_0$ where $\Delta f$ is the spectral half-width at the 1/$e$-intensity point. Equation (\ref{eq:t}) indicates that pulse focusing occurs only when $Ck_2<0$, where an up-chirped $C>0$ (down-chirped $C<0$) pulse is used as an input in an anomalous (normal) dispersion regime $k_2<0$ ($k_2>0$).

As an experimental demonstration in this device, an up-chirped pulse is injected in the anomalous dispersion regime between 3.5-7 MHz as shown in Fig. 4(a). The output pulse width is measured at various distances by exciting different chirped pulses with a centre frequency of 5.8 MHz as shown in Figs. 4(b)-(d). 
The pulse widths with negative $C$ monotonically increase with increasing distance (see Fig. 4(b) and the upper panel of Fig. 4(c)), whereas the widths with a positive $C$ first decrease to the transform limited value, and in turn, increase with distance (see Fig. 4(b) and the lower panel of Fig. 4(c)). 
It should be also noted that a smaller pulse width is observed when we employ a larger absolute value of positive $C$ and indeed, strong focusing is realized when $C = 9.7$ where the pulse width is compressed from 2 $\mu$s to 0.5 $\mu$s and the strain energy is enhanced more than one order of magnitude. 
Making use of the GVD effect enables a travelling pulse waveform to be engineered that leads to the temporal focusing of phonons.

In a conventional spatial lens, the figures of merit such as the compression and amplification factors, and the ability to spatially adjust the focusing position, are mainly determined by the geometric parameters of the system. 
Therefore, these techniques require sophisticated artificial structures if they are to control waves. In contrast this time lens enables them to be dynamically designed by simply changing the excitation frequency, input pulse width and chirp parameters. Although, in this study, these focusing properties are limited by the device structure causing the reflection at both WG edges, and the finite bandwidth of the lock-in amplifier, it is possible to further enhance this focusing effect by modifying the device structure and optimizing the measurement set-up. 

In conclusion, we have demonstrated a phononic time lens in a 1D PnC WG.
The dispersion effect determined by the periodic geometry of the device induces frequency chirp in the Gaussian input pulse during propagation. This phenomenon also allows the travelling wave to be temporally focused, and this can be controlled by changing the input chirp parameters. This novel temporal focusing can enhance the availability of mobile phonons, and this ability will open up potential applications to investigate nonlinear phononic phenomena.
\\\\
{\bf Methods}\\
Nanomechanical vibrations in the PnC WG were excited by applying an amplitude modulated alternating voltage from a signal generator (NF Wavefactory 1974 and 1968), and were measured with a He-Ne laser Doppler interferometer (NEOARK MLD-230V-200-NN). In the spectral measurements in Fig. 1(b), the electrical output from the interferometer was measured with a vector signal analyzer (HP89410A). In the temporal measurements in Figs. 2(a), 3(b)-(c) and 4(b)-(c), the electrical output was first filtered by a lock-in amplifier (Zurich Instruments HF2LI), and then measured with an oscilloscope (Agilent DSO6014A). 
The spectral bandwidth of the low-pass filter in the lock-in amplifier was set at 470 kHz, which is equivalent a time constant of 102.6 ns.\\\\
{\bf Acknowlegements}\\
We are grateful to Y. Ishikawa for growing the heterostructure. This work is partly supported by a MEXT Grant-in-Aid for Scientific Research on Innovative Areas "Science of hybrid quantum systems" (Grant No. JP15H05869).\\\\
{\bf Author Contributions}\\
M.K. and D.H. performed the measurements and data analysis. D.H. fabricated the sample and K.O. co-fabricated the GaAs/AlGaAs heterostructure. M.K. and D.H. wrote the paper and H.Y. planned the project.

\newpage

\begin{figure*}[!hbtp]
\begin{center}
\vspace{0cm}\hspace{-0cm}
\includegraphics[scale=0.13]{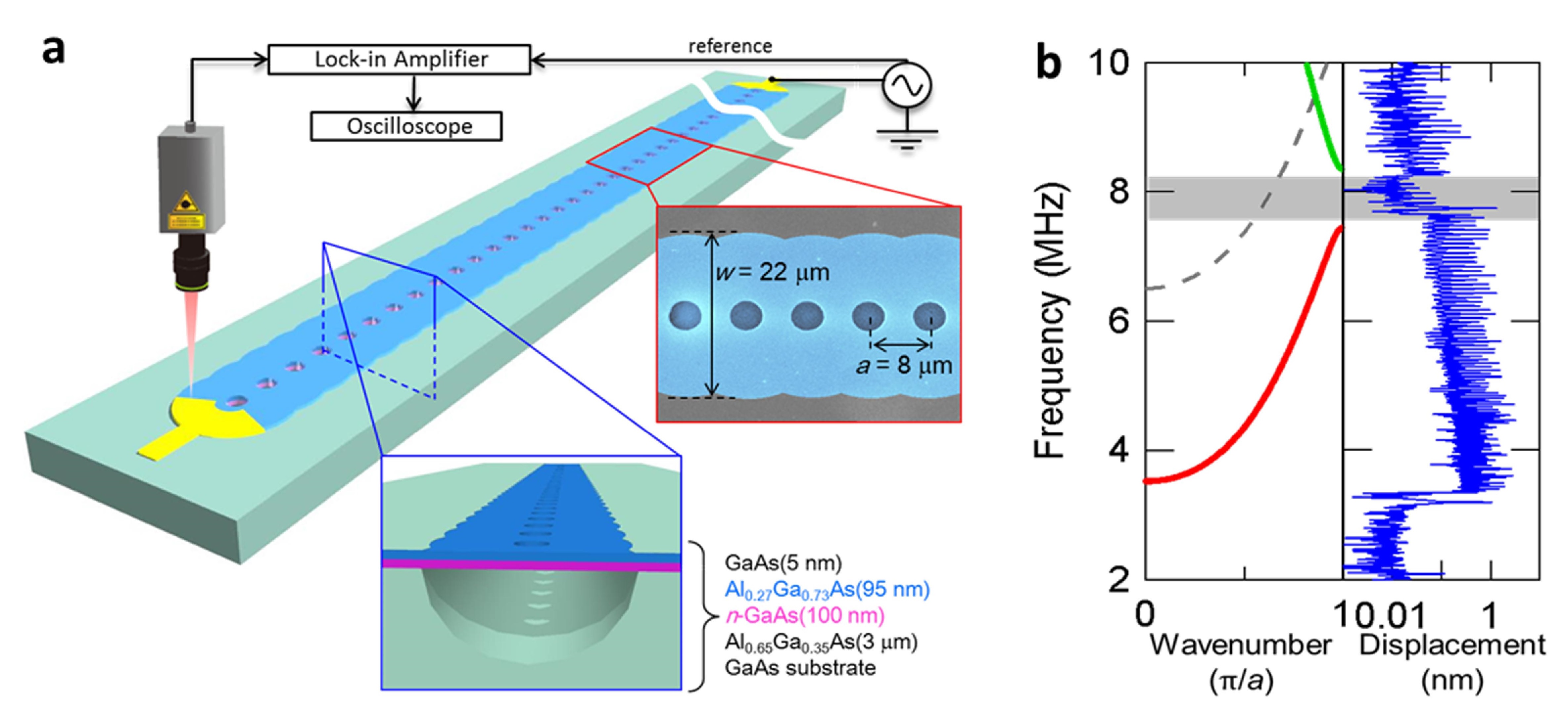}
\vspace{-0cm}
\caption{{\bf 1D phononic crystal waveguide.} ({\bf a}) A schematic of the PnC WG and the measurement set-up. The device has a GaAs/AlGaAs heterostructure fabricated by selectively etching the Al$_{0.65}$Ga$_{0.35}$As layer as shown in the bottom inset, and the periodic structures are determined by the WG width of 22 $\mu$m and the hole pitch of 8 $\mu$m as shown in the right inset. The nanomechanical vibrations are piezoelectrically excited at the right edge and detected at the left edge with a laser Doppler interferometer at room temperature and in a high vacuum (2 $\times 10^{-4}$ Pa). ({\bf b}) The FEM simulated dispersion relation of the PnC WG (left panel) and the experimental transmission spectrum, which is measured at the left edge by employing continuous excitation with 1.0 V$_{\rm rms}$ from the right edge (right panel). 
The internal stress existing between the GaAs and Al$_{0.27}$Ga$_{0.73}$As layers is included in the simulation\cite{Liu_2011}. The dashed line indicates the 2nd phonon band, which does not contribute to the transmission due to the piezoelectric transducer electrodes being located at the nodal position of the mode in this band. }
\label{fig 1}
\end{center}
\end{figure*}

\begin{figure}[!t]
\begin{center}
\vspace{0cm}\hspace{0cm}
\includegraphics[scale=0.12]{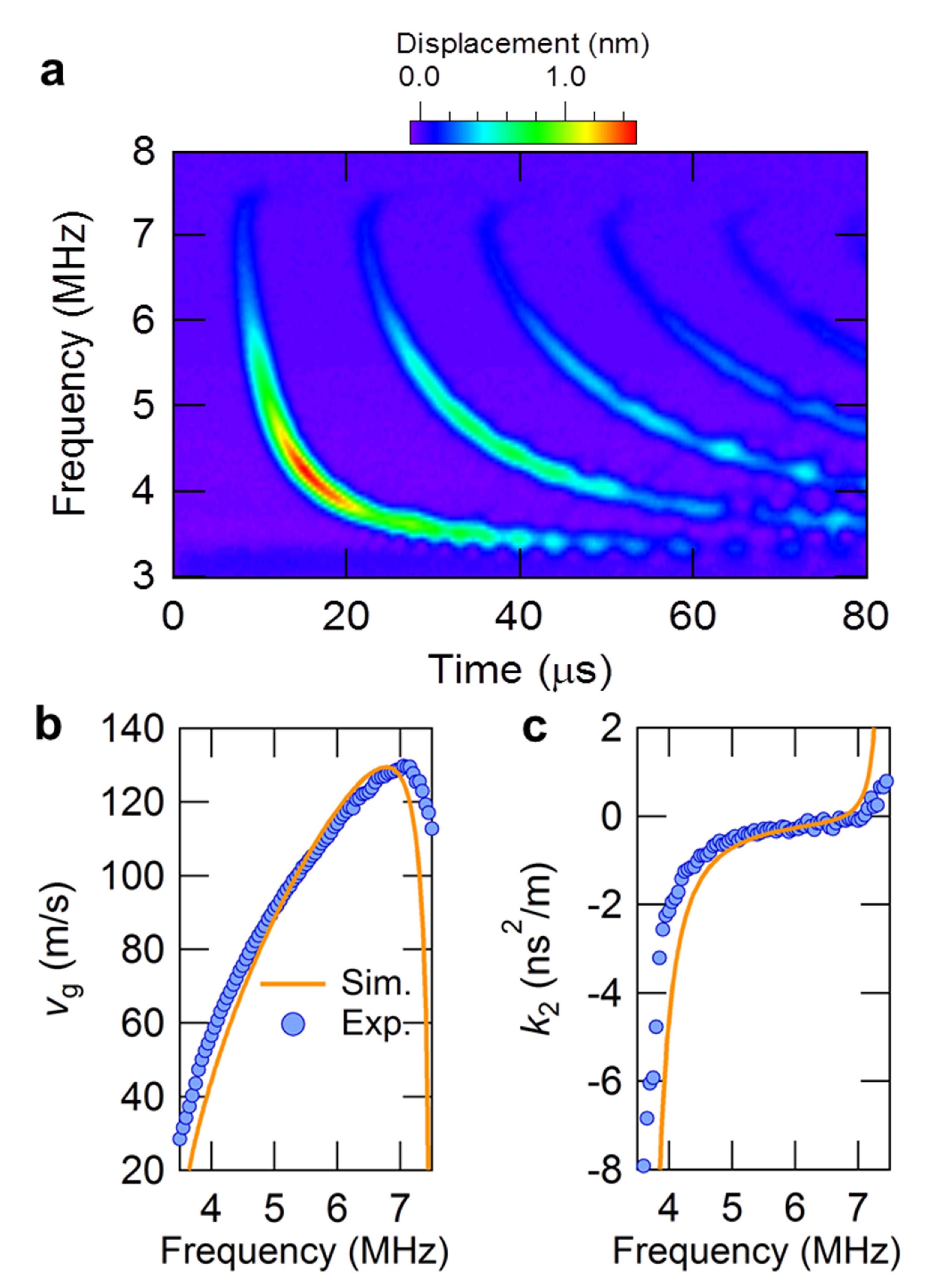}
\vspace{-0cm}
\caption{{\bf Fundamental properties of the PnC WG.} 
({\bf a}) The temporal response of phonon propagations measured at the left edge of the WG when exciting the input Gaussian pulse with $T_0=1.2\ \mu$s and an amplitude of 1.0 V$_{\rm rms}$ from the right edge.
({\bf b}), ({\bf c}) The frequency dependence of the group velocity $v_{\rm g}$ and the GVD coefficient $k_2$ respectively, where the experimental and FEM simulated results are denoted as circles and solid lines.
}
\label{}
\end{center}
\end{figure}

\begin{figure*}[!t]
\begin{center}
\vspace{0cm}\hspace{0cm}
\includegraphics[scale=0.8]{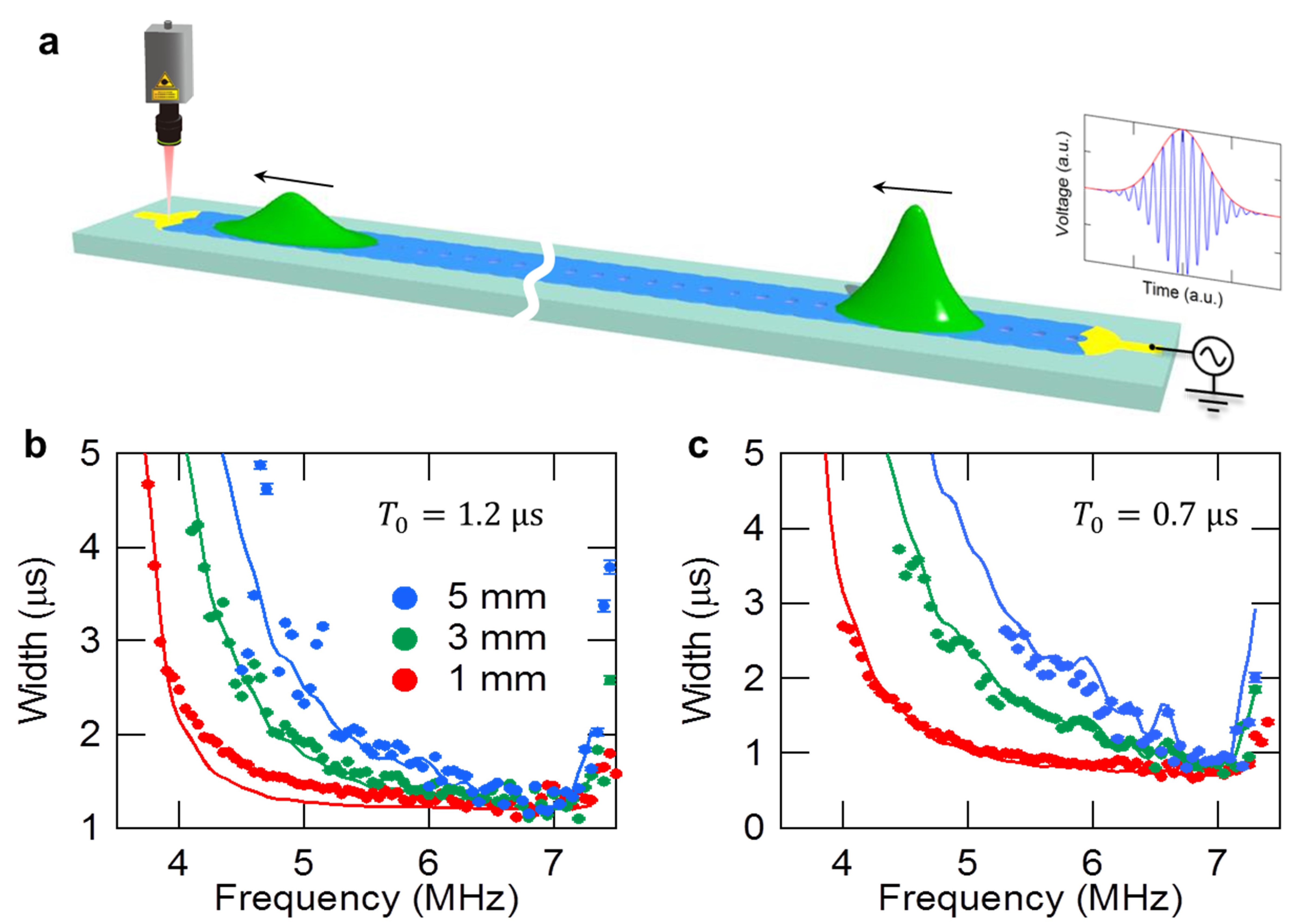}
\vspace{-0cm}
\caption{{\bf Pulse broadening due to the group velocity dispersion effect.} 
({\bf a}) A schematic showing the pulse broadening due to the GVD effect when injecting an un-chirped Gaussian pulse.
({\bf b}), ({\bf c}) The frequency dependence of the temporal width in the output waveform measured at the propagation distance $x=$ 1 (red), 3 (green) and 5 (blue) mm by exciting Gaussian input pulses with $T_0=1.2$ $\mu$s and 0.7 $\mu$s, respectively.
The solid line indicates the theoretical results estimated by substituting the experimentally-obtained GVD coefficient $k_2$ in Fig. 2({\bf c}) into equation (\ref{eq:uuu}).
}
\end{center}
\end{figure*}

\begin{figure*}[!t]
\begin{center}
\vspace{0cm}\hspace{0cm}
\includegraphics[scale=0.8]{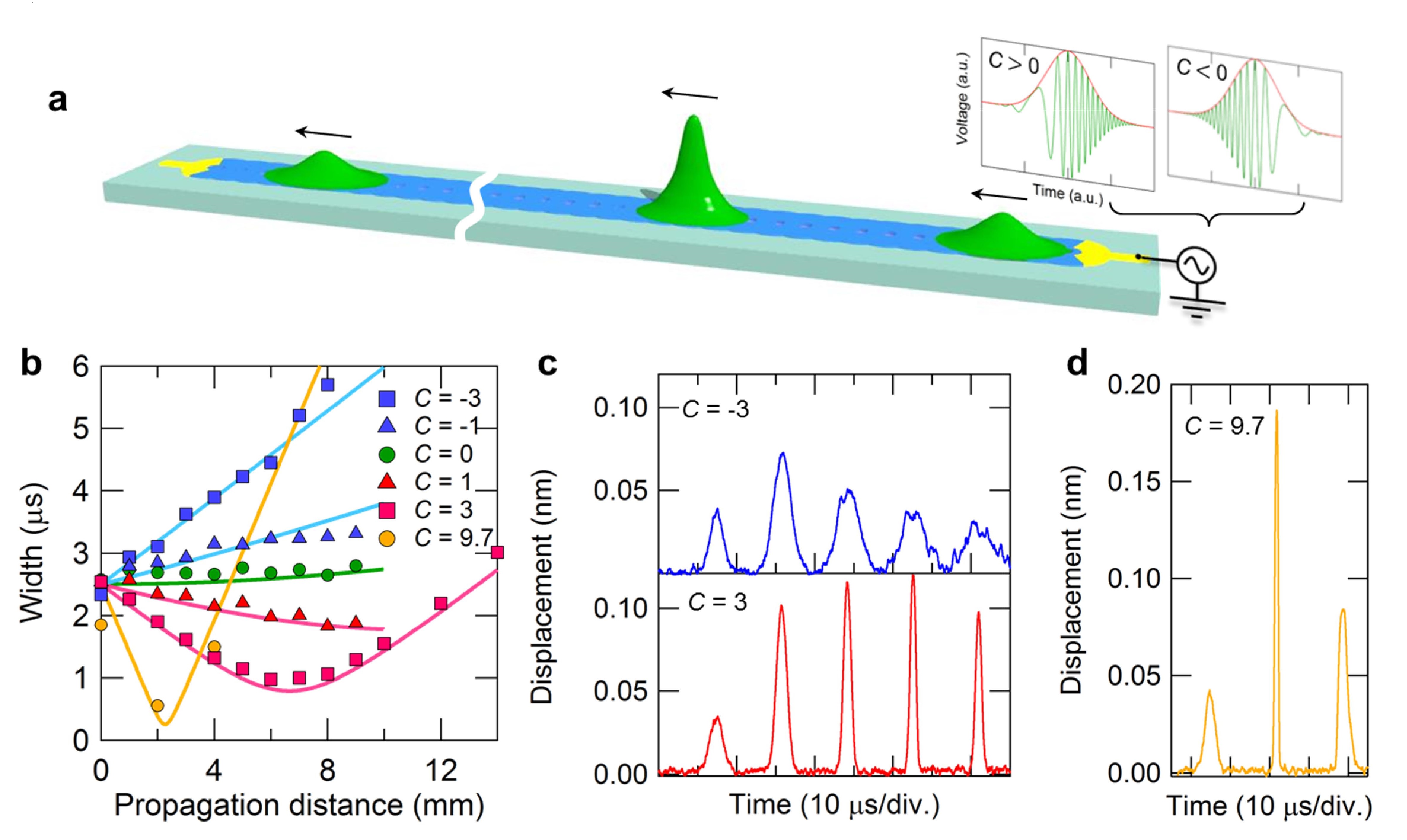}
\vspace{-0cm}
\caption{{\bf Temporal focusing of phonon pulses.} ({\bf a}) A schematic showing the pulse focusing caused by the GVD effect when injecting a chirped Gaussian pulse. ({\bf b}) Temporal pulse width as a function of propagation distance when exciting a 2.5 $\mu$s  chirped pulse with a centre frequency of 5.8 MHz and various chirp parameters. The GVD coefficient $k_2$ at this frequency is -0.28 ns$^{2}$/m in Fig. 2(c), indicating an anomalous dispersion regime. ({\bf c}), ({\bf d}) Temporal evolution of the phonon wave when exciting a chirped Gaussian pulse with a chirp parameter $C=\pm 3$ and 9.7 and measuring it at the right edge, respectively. The observed wave packets except for the first one are associated waves composed of the incident and reflected waves at the WG edge, which increase the peak amplitude nearly twofold while maintaining the pulse width.}
\end{center}
\end{figure*}

\clearpage

\renewcommand{\thesection}{S\arabic{section}}
\renewcommand{\thesubsection}{S\arabic{section}.\arabic{subsection}}
\renewcommand{\figurename}{\small{Figure S}\hspace{-0.11cm}}
\renewcommand{\theequation}{S\arabic{equation}}
\setcounter{figure}{0}

\begin{center}

\noindent \textbf{{\Large Supplementary Information}}\\

\vspace{0.3cm}

\noindent{{\Large On-chip phononic time lens}}\\

\vspace{0.3cm}

\noindent \small{{\large M. Kurosu$^{1,2}$ D. Hatanaka$^{1}$, K. Onomitsu$^{1}$ and H. Yamaguchi$^{1,2}$}}\\
\vspace{0.2cm}
\noindent \small{\large\it $^1$NTT Basic Research Laboratories, NTT Corporation, Atsugi-shi, Kanagawa 243-0198, Japan}\\
\noindent \small{\large\it $^2$Department of Physics, Tohoku University, Sendai 980-8578, Japan}\\
\end{center}

\vspace{0.5cm}

\section{Effect of third order dispersion}

The GVD effect, which is proportional to $k_2$, determines the phonon pulse waveform. However, we also need to take account of the third-order dispersion (TOD) effect $k_3=\frac{\partial^3k}{\partial\omega^3}$ when the centre frequency of the pulse approaches the band edges (ref. S1). To evaluate the contribution of these effects, it is useful to introduce dispersion length scales for GVD and TOD (ref. S2),
\begin{equation}
x_D = \frac{T_0^2}{|k_2|},\ \ \ x_D^\prime = \frac{T_0^3}{|k_3|} \tag{S1},
\label{eq:ld}
\end{equation} 
Using equation (S1), equation (6) in the main text can be modified to,
\begin{equation}
{\rm i}\frac{\partial U}{\partial x} = \frac{k_2}{2}\frac{\partial ^2 U}{\partial T^2}+\frac{i k_3}{6}\frac{\partial ^3 U}{\partial T^3}=\frac{{\rm sgn}(k_2)T_0^2}{2x_D}\frac{\partial ^2 U}{\partial T^2}+i\frac{{\rm sgn}(k_3)T_0^3}{6x_D^\prime}\frac{\partial ^3 U}{\partial T^3} \tag{S2}
\label{}
\end{equation} 
where ${\rm sgn}(k_{i=2,3}) =\pm1$ depending on the signs of $k_2$ and $k_3$. Equation (S2) indicates that the GVD effect is dominant in pulse evolution when $x_D$ is smaller than $x_D^\prime$, but the TOD effect also becomes dominant when $x^\prime_D$ approaches $x_D$. 
Supplementary Figure 1(a)-(c) shows the frequency dependence of $x_D, x^\prime_D$ and the ratio $x^\prime_D/x_D$ of 1D PnC WG, respectively, which are calculated from an FEM simulation. These results indicate that the pulse evolution is mainly dominated by the GVD effect around the centre of the band, where $x^\prime_D/x_D$ is sufficiently large, and thus, the temporal focusing of the phonon wave at 5.8, 5.35 and 4.5 MHz can be described solely by the GVD effect as shown in Fig. 4(b), Supplementary Fig. 2(a) and 2(b), respectively. On the other hand, $x^\prime_D/x_D$ is small at the band edges, which allows the TOD effect to distort the pulse waveform, where it becomes asymmetric and an oscillatory structure appears near the tailing edge as shown in Supplementary Fig. 3(a), and thus the dynamics of the temporal focusing deviates from the GVD theory as shown in Supplementary Fig. 3(b).\\\\

\begin{figure}[!htb]
\begin{center}
\vspace{0cm}\hspace{0cm}
\includegraphics[scale=0.12]{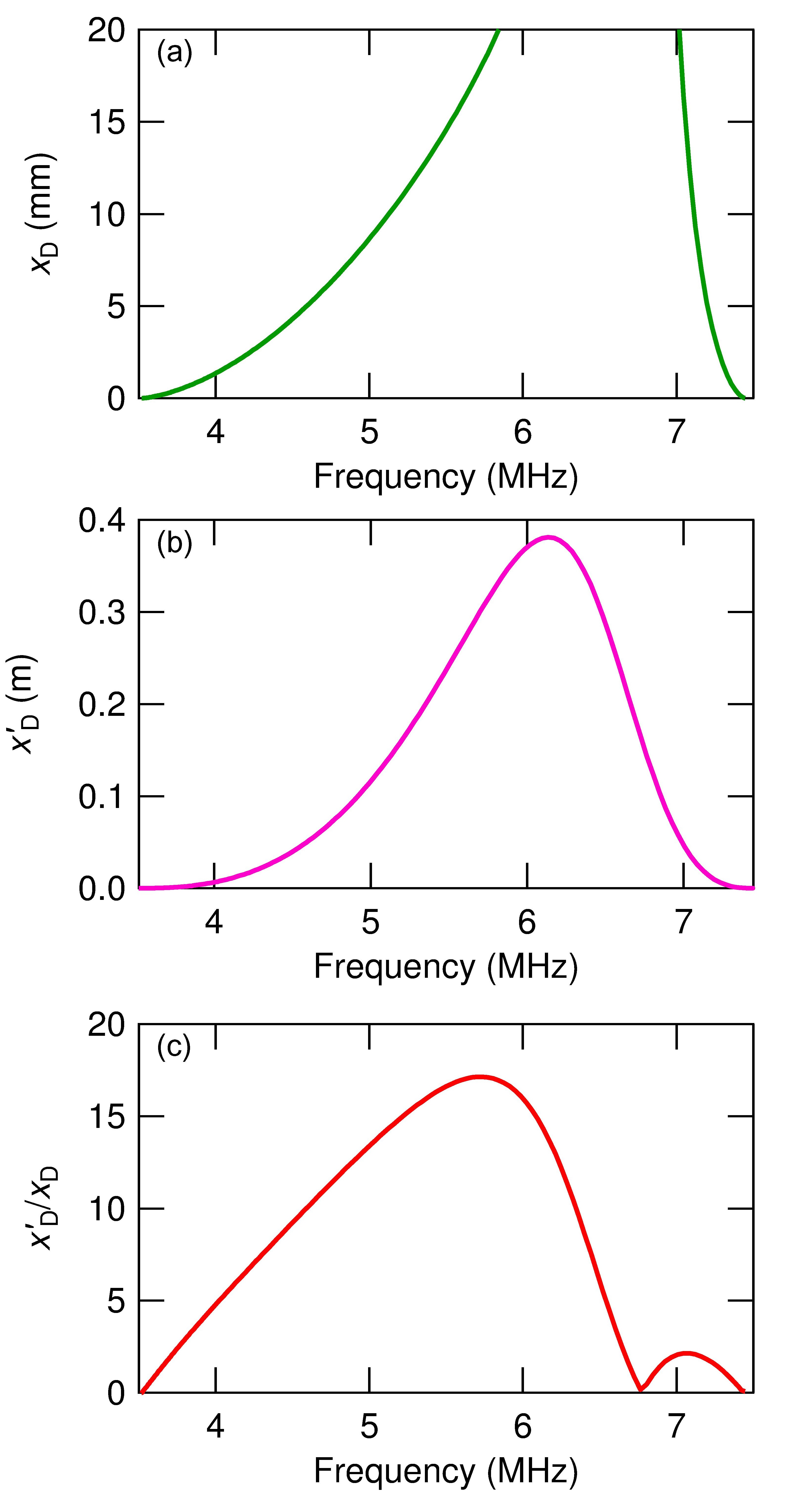}
\caption{{\bf Higher order dispersion length.} 
({\bf a})-({\bf c}) The frequency dependence of $x_D$, $x_D^\prime$ and the ratio of $x_D^\prime$ to $x_D$, respectively. These values are calculated from equation (\ref{eq:ld}) using $T_0 = 2.5\ \mu$s.
}
\end{center}
\end{figure}
\begin{figure}[!htb]
\begin{center}
\vspace{0cm}\hspace{0cm}
\includegraphics[scale=0.13]{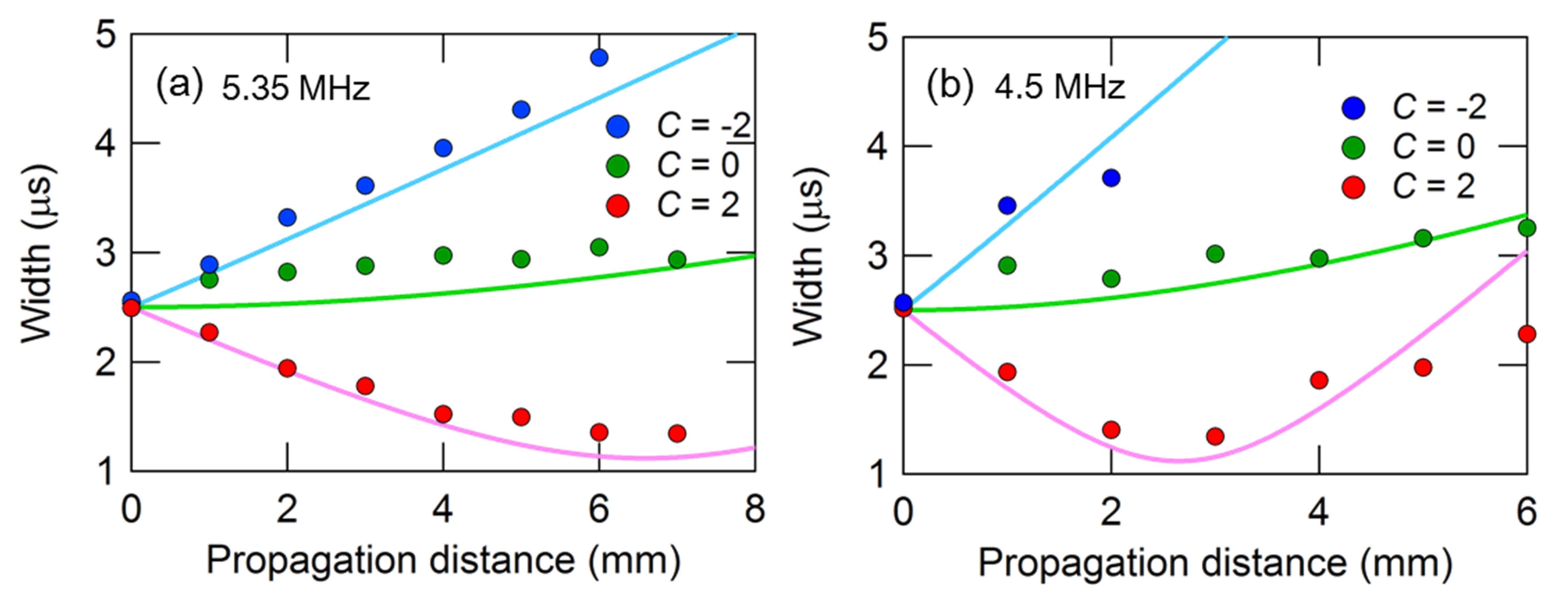}
\caption{{\bf Second order dispersion effect.} 
({\bf a}), ({\bf b}) Temporal pulse width as a function of propagation distances when exciting a 2.5 $\mu$s chirped pulse with the centre frequencies of 5.35 and 4.50 MHz with chirp parameters of $\pm 2$ and 0 respectively. Solid lines indicate the theoretical results when $x_D^\prime/x_D=15.8$ (a) and 9.2 (b).
}
\label{Supplementary Figure 1}
\end{center}
\end{figure}
\begin{figure}[!htb]
\begin{center}
\vspace{0cm}\hspace{0cm}
\includegraphics[scale=0.13]{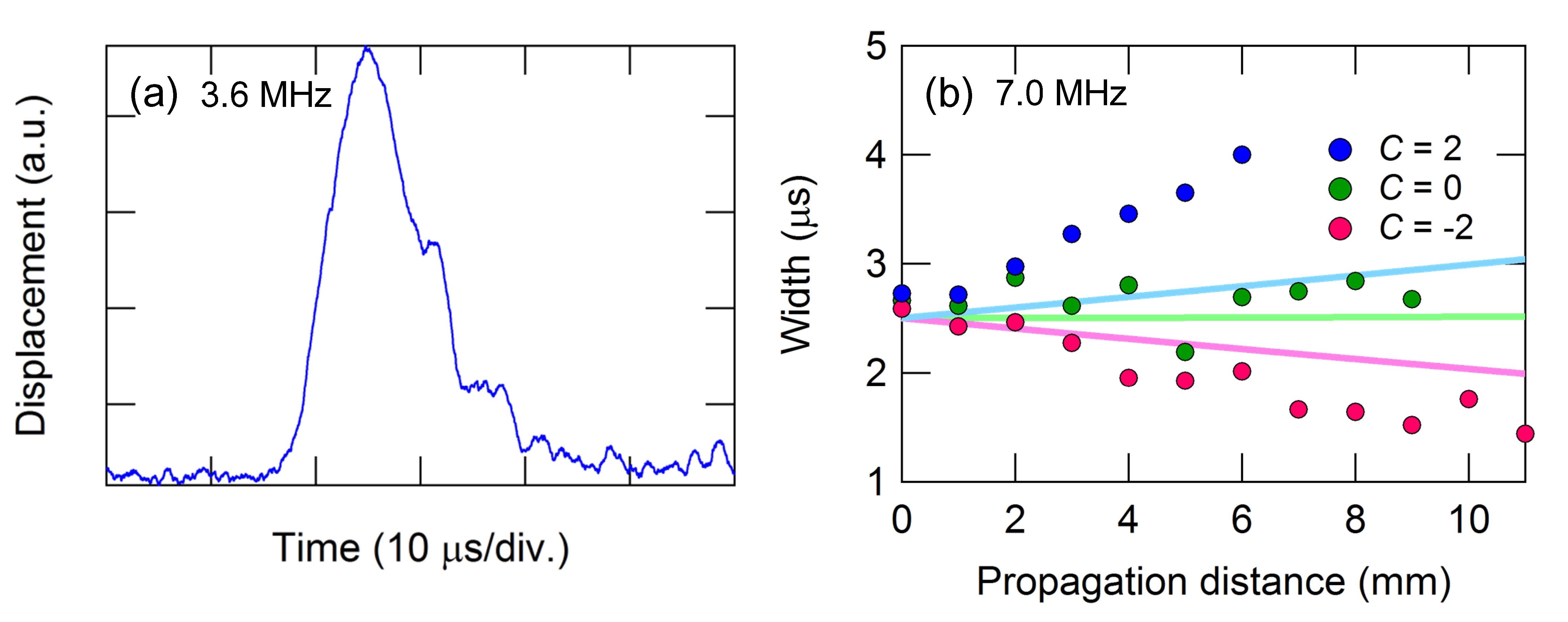}
\caption{{\bf Third order dispersion effect.} 
({\bf a}) Temporal response of a phonon wave by exciting the un-chirped Gaussian input pulse with $T_0=1.2$ $\mu$s when $x_D^\prime/x_D=0.8$.
({\bf b}) Temporal pulse width as a function of propagation distances when exciting a 2.5 $\mu$s chirped pulse with a centre frequency of 5.8 MHz with chirp parameters of $\pm 2$ and 0. Solid lines indicate the theoretical results when $x_D^\prime/x_D=2.1$.
}
\label{Supplementary Figure 1}
\end{center}
\end{figure}

\clearpage
\noindent {\bf {\Large References}}\\

\noindent S1. Marcuse, D. Pulse distortion in single-mode fibers. 3: Chirped pulses. $Appl. Optics$ {\bf 20}, 

\ \ 3573 (1981).

\noindent S2. Agrawal, G. $Nonlinear Fiber Optics$. Boston, fifth edition, (2013).

\end{document}